\documentclass[final,5p,times,twocolumn]{elsarticle} 
\usepackage{graphicx}
\usepackage{amsmath}
\usepackage{mathtools}
\usepackage{amssymb}
\usepackage{lineno}
\usepackage[colorinlistoftodos]{todonotes}
\usepackage[colorlinks=true, allcolors=blue]{hyperref}
\journal{Nuclear Instruments and Methods A}

\usepackage{multirow}
\usepackage{multicol}
\usepackage{comment} 

\begin{document}
\begin{frontmatter}

\title{Radiation damage uniformity in a SiPM}
\author[B]{O.~Bychkova\corref{mycorrespondingauthor}}
\cortext[mycorrespondingauthor]{Corresponding author}
\ead{oksana.bychkova@cern.ch}
\author[A]{E.~Garutti}
\author[B]{E.~Popova} 
\author[B]{A.~Stifutkin}
\author[A]{S.~Martens}
\author[B]{P.~Parygin}
\author[C]{A.~Kaminsky}
\author[A]{J.~Schwandt}

\address[A]{University of Hamburg, 22761, Luruper Chaussee 149, Hamburg, Germany}
\address[B]{National Research Nuclear University MEPhI (Moscow Engineering Physics Institute), 115409, Kashirskoe Shosse 31, Moscow, Russia}
\address[C]{Skobeltsyn Institute of Nuclear Physics, Lomonosov Moscow State University, 119991, 1(2), Leninskie Gory, GSP-1, Moscow, Russia}

\date{\today}
\graphicspath{ {fig/} }

\begin{abstract}
A dedicated single-cell SiPM structure is designed and measured to investigate the radiation damage effects on the gain and breakdown voltage of SiPMs exposed to a reactor neutron fluence up to $\Phi$~=~5e13~cm$^{-2}$. The cell has a pitch of 15~$\mu$m. Results of the measurements and analysis of the IV-curves are presented. Impact of the self-heating effect was investigated. The radiation damage uniformity of 1 cell and 120 cells was checked up to $U_\mathit{ov}$~=~1.7~V. Fluence dependence of the breakdown voltage from the current measurements $U^{IV}_\mathit{bd}$ was extracted and compared to that of the breakdown voltage from the gain measurements $U^{G}_\mathit{bd}$.
\end{abstract}

\begin{keyword}
Silicon photomultiplier
\sep
radiation damage
\sep
single cell SiPM 
\end{keyword}
\end{frontmatter}

\section{Introduction}
Silicon photomultipliers (SiPMs)~\cite{Piemonte_2019}, thanks to their
excellent performance, are becoming the photodetectors of choice
for many applications. One major limitation, in particular for
their use at high-luminosity colliders, is the radiation damage induced by charged or neutral hadrons. 
As SiPMs detect single charge carriers, radiation damage is a major concern when 
operating these devices in harsh radiation environments (i.e. CMS and LHCb detectors at LHC, detectors at the proposed International Linear Collider (ILC), detectors for space experiments, etc.). 
Results on the operation of irradiated SiPMs with X-ray, gamma, electron, proton and neutron sources are reviewed in~\cite{Garutti_2019}. The most critical effect of radiation on SiPMs is the increase of dark count rate, which makes it impossible to resolve signals generated by a single photon from the noise. Once the single photo-electron (SPE) resolution is lost the SiPM gain cannot be directly determined as the separation of the peaks in a SPE distribution. 
Additionally, 
the breakdown voltage $U^{G}_\mathit{bd}$ of the SiPM cannot be determined using the widely applied method of linear dependence of the gain versus bias voltage.
 It should be noted that differences in the values extracted with the two methods have been reported, for instance in Ref.~\cite{Chmill:2016msk}. For the understanding of the possible origin of this difference we refer to~\cite{Marinov2007} and to recent simulation studies in~\cite{cazimajou}. 
 
 We have presented the first results of radiation hardness study using SiPMs with a single-cell
readout in~\cite{BYCHKOVA2022166533}, where the fluence dependence of gain and turn-off voltage $U^{G}_\mathit{bd}$ are investigated.
A reduction of the gain by 19\% and an increase of $U^{G}_\mathit{bd}$ by $\approx$0.5~V is observed after $\Phi$~=~5e13 cm$^{-2}$.
Three outstanding questions related to these studies are addressed in this paper: 
\begin{enumerate}
    \item Are the measurements affected by self-heating effect? 
    \item Is the increase of $U^{G}_\mathit{bd}$ with fluence correlated to that of $U^{IV}_\mathit{bd}$?
    \item Is the radiation damage of a single cell representative of the average radiation damage of the entire SiPM? 
\end{enumerate}

In particular this last point is essential to confirm the validity of the results obtained with the single cell and extend them to the whole SiPM. 

In this paper we present the answers to these questions, analyzing the current-voltage curves of a single cell and of its 120 surrounding cells in a SiPM, measured on sensors irradiated with reactor neutron fluence up to $\Phi$~=~5e13~cm$^{-2}$.

\section{Device and setup description}
\label{sec:setup}
The device under test (DUT) is a Hamamatsu SiPM test structure of S14160 series \cite{HPKS14160} glued on the ceramic package. It consists of one single cell surrounded by 120 others. This single cell can be biased and read out separately. 

\begin{figure}[ht]
    \centering
    \includegraphics[width=0.4\textwidth]{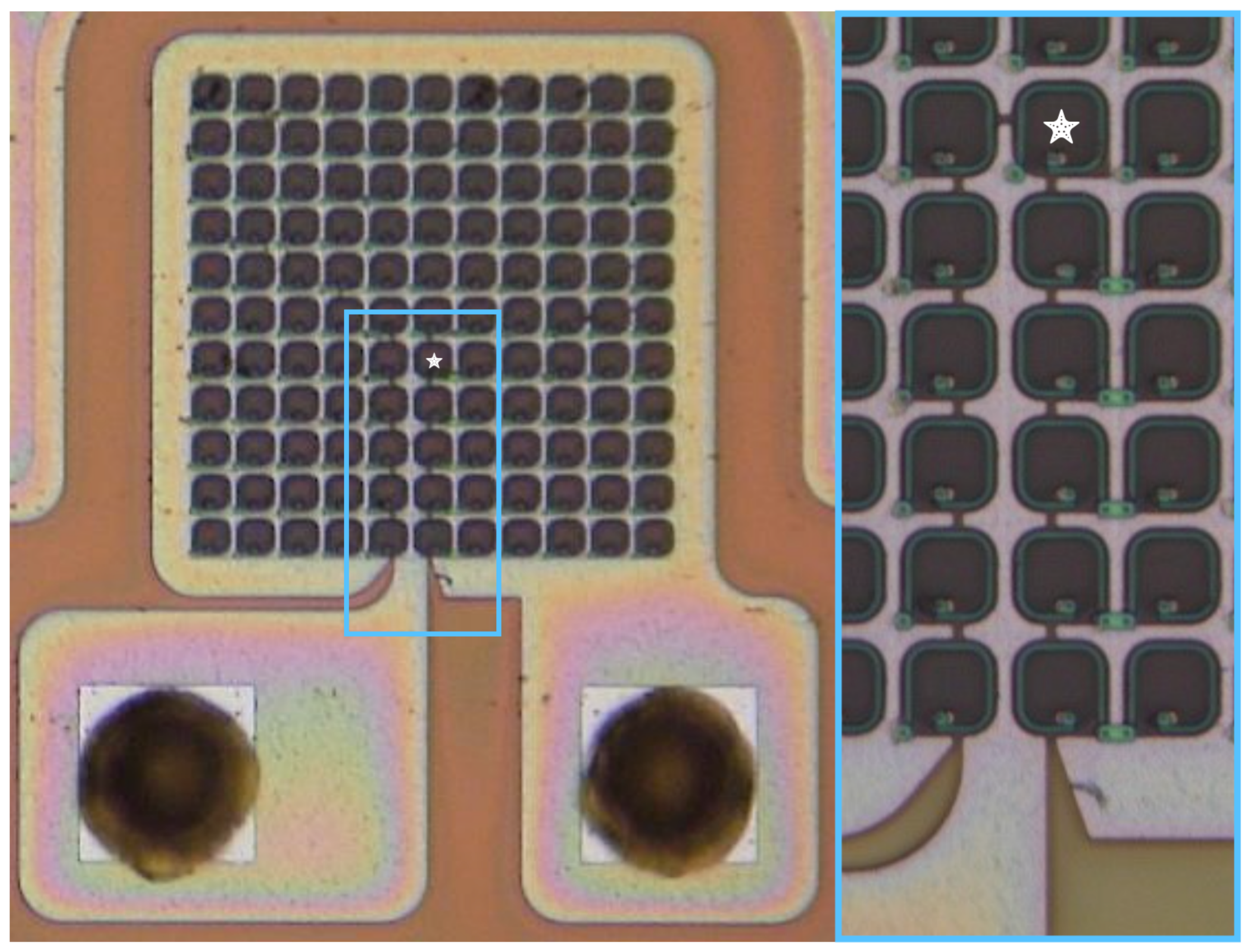}
    \caption{Microscopic image of the studied 11x11-cell array. The disconnected cell is marked with a star symbol. On the 50x-zoomed image on the right-hand side one can see the cell isolation by gaps in the metallization layer.}
    \label{fig:DUTpic2}
\end{figure}

A picture of the DUT is shown in Figure~\ref{fig:DUTpic2}. It has an array of 11x11 cells with 15~$\mu$m pitch. The central cell of the array is disconnected from the others and has its own output contact pad. Therefore 1 cell and 120 cells have a common cathode but separate anodes. 
Between the cells, trenches of 0.5~$\mu$m width are implemented to reduce optical cross-talk \cite{hpkCHEF2019}. 

The setup for IV-measurements consists of a climate chamber, a dual-channel bias and readout board, SourceMeter Keithley 2450 used for 1 cell and SourceMeter Keithley 6517b used for 120 cells. To monitor the temperature as close to the DUT as possible, a Pt-100 is attached to the side of the ceramic package. The temperature stability in the chamber monitored by the Pt-100 readings is $\pm$0.03~$^{\circ}$C and $\pm$0.2~$^{\circ}$C at +20~$^{\circ}$C and -30~$^{\circ}$C, respectively. The accuracy of the setup on low current measurements is limited, such that currents lower than 100~pA for Keithley 2450 and 1~nA for Keithley 6517b are not reliable.  

For the illumination of the device a stabilised broadband light source with a 650~nm filter is used. A filter mount placed outside the chamber is used for changing the light intensity. The light delivery system consists of two optical fibers, the first with 365~$\mu$m-core and 1~m length from the light source to the filter mount, and the second with 365~$\mu$m-core and 20~m length from the filter mount to the DUT. For the measurements without illumination the optical fiber was blocked in the filter mount by a beam blocker, while the light source was kept on. 

\begin{figure}[ht]
    \centering
    \includegraphics[width=0.7\linewidth]{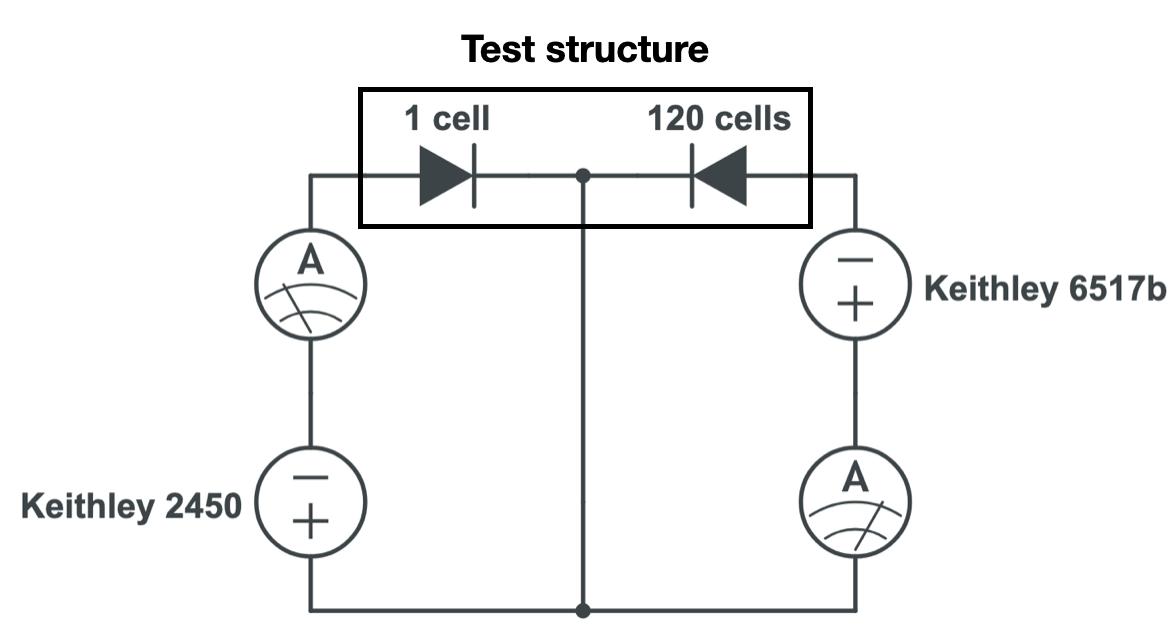}
    \caption{Circuit schematic of the bias and readout board for IV-measurements.}
    \label{fig:IVboard}
\end{figure}

The readout board consists of two channels to bias and read out 1 cell and 120 cells separately. The circuit schematic of the board is shown in Figure \ref{fig:IVboard}. IV-curves are measured synchronously on 1 cell and 120 cells. 

Measurements were carried out for one non-irradiated device and three devices irradiated by neutrons at the TRIGA Research Reactor of the JSI, Ljubljana, to different fluences $\Phi$ = [2e12, 1e13, 5e13]~cm$^{-2}$. No annealing was applied to the samples before measurement.
\begin{comment}
The neutron irradiations were performed at room temperature without applied bias at the TRIGA Research Reactor of the JSI, Ljubljana. The samples were transported cold to Hamburg after irradiation and stored in a refrigerator
at -30~$^{\circ}$C. No annealing was applied to the samples before measurement.
\end{comment}

\section{Self-heating}
The highest heat power observed in our study was $P_\mathit{heat}$ = 1.9~mW for 120 cells irradiated to $\Phi$~=~5e13~cm$^{-2}$, at T~=~-30~ $^{\circ}$C and $U_\mathit{ov}$~=~4~V. In this paper the overvoltage is defined as $U_\mathit{ov} = U_\mathit{bias} - U^{IV}_\mathit{bd}$.
This power dissipated inside the SiPM could lead to a local increase of temperature and a correlated change of SiPM performance parameters, an effect denoted as self-heating. 
To check whether the measurements are affected by the self-heating effect the following procedure was carried out for the non-irradiated sample at T~=~+20~$^{\circ}$C:
\begin{itemize}
    \item Operate 120 cells at the fixed voltage above the breakdown ($U_\mathit{bias}$~=~39.1~V, $U_\mathit{ov}$~=~1.5~V). 
    \item Change the light intensity to control the heat power, thus the 120 cells serve as a heater. 
    \item Measure the IV-curve of 1 cell and calculate $U^{IV}_\mathit{bd}$, thus the 1 cell serves as a temperature sensor since $U_\mathit{bd}$ strongly depends on the SiPM temperature. 
\end{itemize}

\begin{figure}[ht]
    \centering
    \includegraphics[width=0.9\linewidth]{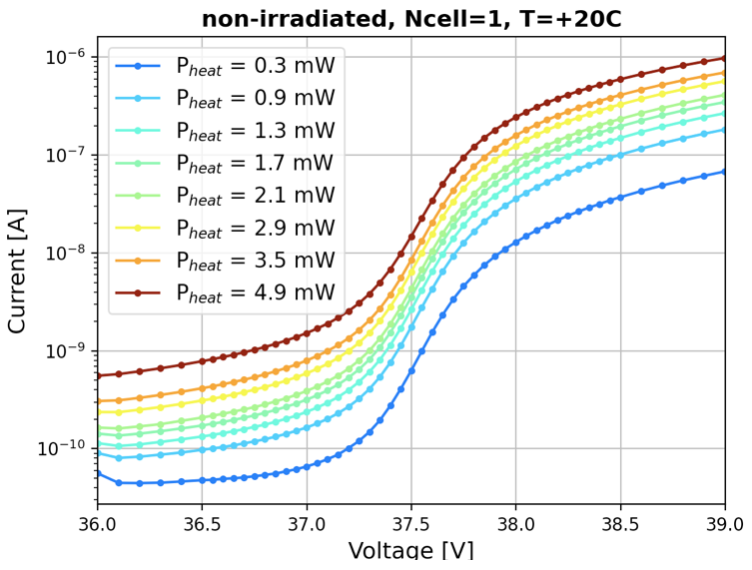}
    \caption{IV-curves of the non-irradiated single cell measured with different light intensity and heat power generated by 120 cells operated at a fixed bias voltage $U_\mathit{bias}$~=~39.1~V ($U_\mathit{ov}$~=~1.5~V).}
    \label{fig:IV_selfheat}
\end{figure}

\begin{figure}[ht]
    \centering
    \includegraphics[width=0.9\linewidth]{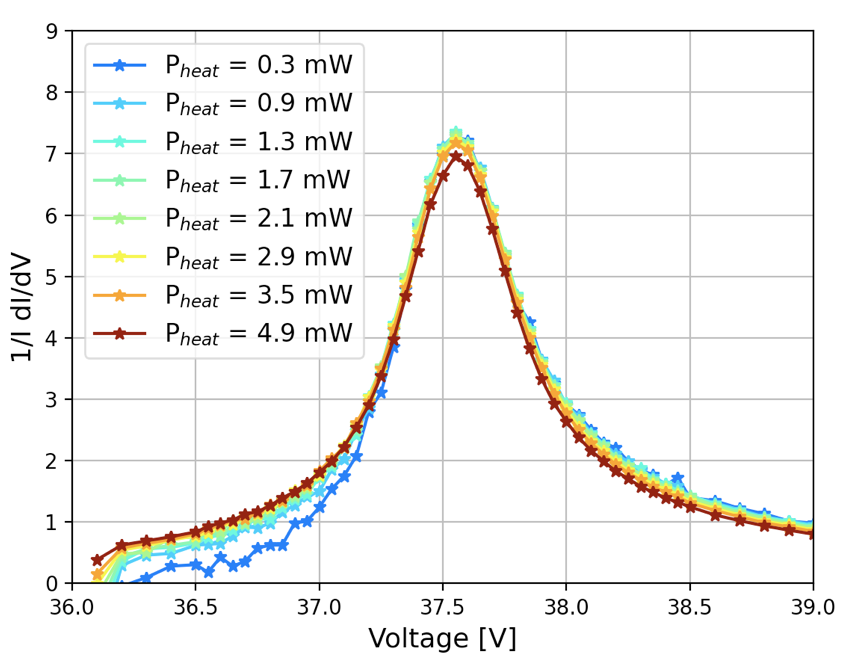}
    \caption{Logarithmic derivative of the IV-curves of the non-irradiated single cell measured with different heat power generated by neighbour 120 cells.}
    \label{fig:Vbd_selfheat}
\end{figure}

Figure~\ref{fig:IV_selfheat} shows the IV-curves for 1 cell, measured with different heat power generated by the other 120 cells. Using the Logarithmic Derivative method \cite{Klanner_2019}, $U_\mathit{bd}$ is calculated for each IV-curve (see Figure~\ref{fig:Vbd_selfheat}). We conclude that the self-heating effect is negligible in our study, since no shift in $U^{IV}_\mathit{bd}$ of a single cell is observed up to $P_\mathit{heat}$~=~4.9~mW, equivalent to the highest power measured in the highest irradiated sensor. 

%\section{Measurements}

\section{Results}
\begin{figure}[h]
  \includegraphics[width=0.9\linewidth]{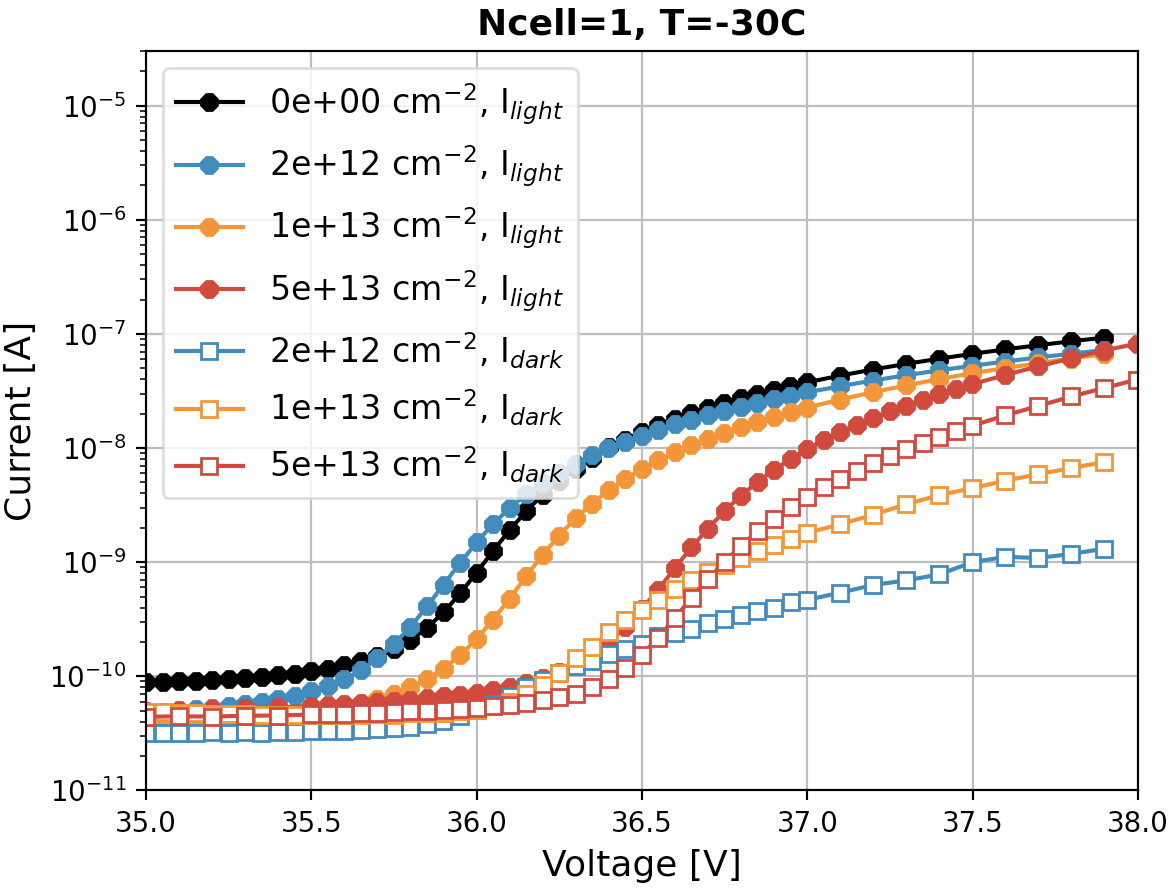}
  \includegraphics[width=0.9\linewidth]{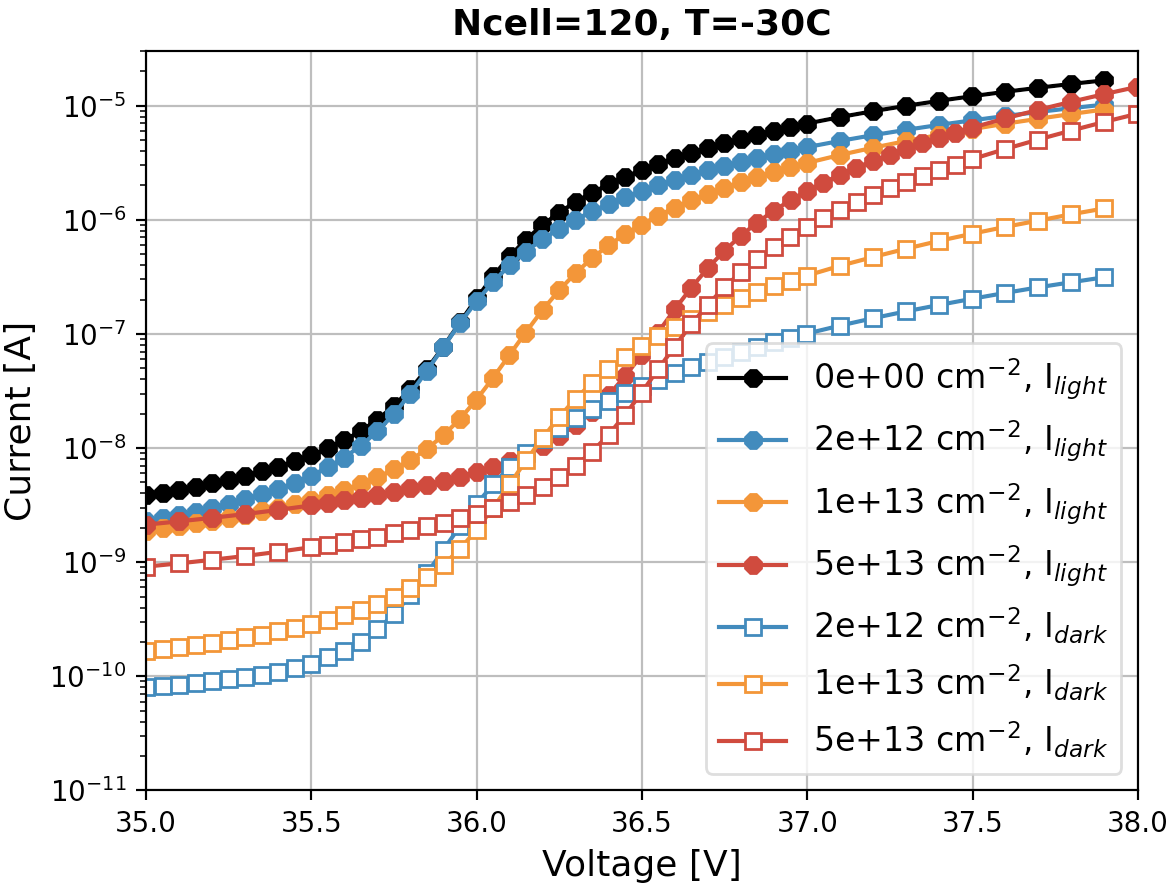}
  \centering
  \caption{IV-curves of a single cell (top) and 120 cells (bottom) for the devices measured at T~=~-30~$^{\circ}$C. Data for non-irradiated (black), $\Phi$~=~2e12~cm$^{-2}$ (blue), $\Phi$~=~1e13~cm$^{-2}$ (orange) and $\Phi$~=~5e13~cm$^{-2}$ (red) samples are shown.}
  \label{fig:IVs}
\end{figure}

Measurements of the IV-curves for 1 cell and 120 cells were carried out for each device without illumination ($I_\mathit{dark}$) and with illumination ($I_\mathit{light}$), at the temperature T~=~-30~$^{\circ}$C in a voltage range $U^{IV}_\mathit{bd}$ - 2~V $< U_\mathit{bias} < U^{IV}_\mathit{bd}$ + 2~V. IV-curves are presented in Figure~\ref{fig:IVs}. Photocurrent $I_\mathit{photo}$ was calculated as a difference of $I_\mathit{light}$ and $I_\mathit{dark}$.

\begin{figure}[ht]
    \centering
    \includegraphics[width=0.9\linewidth]{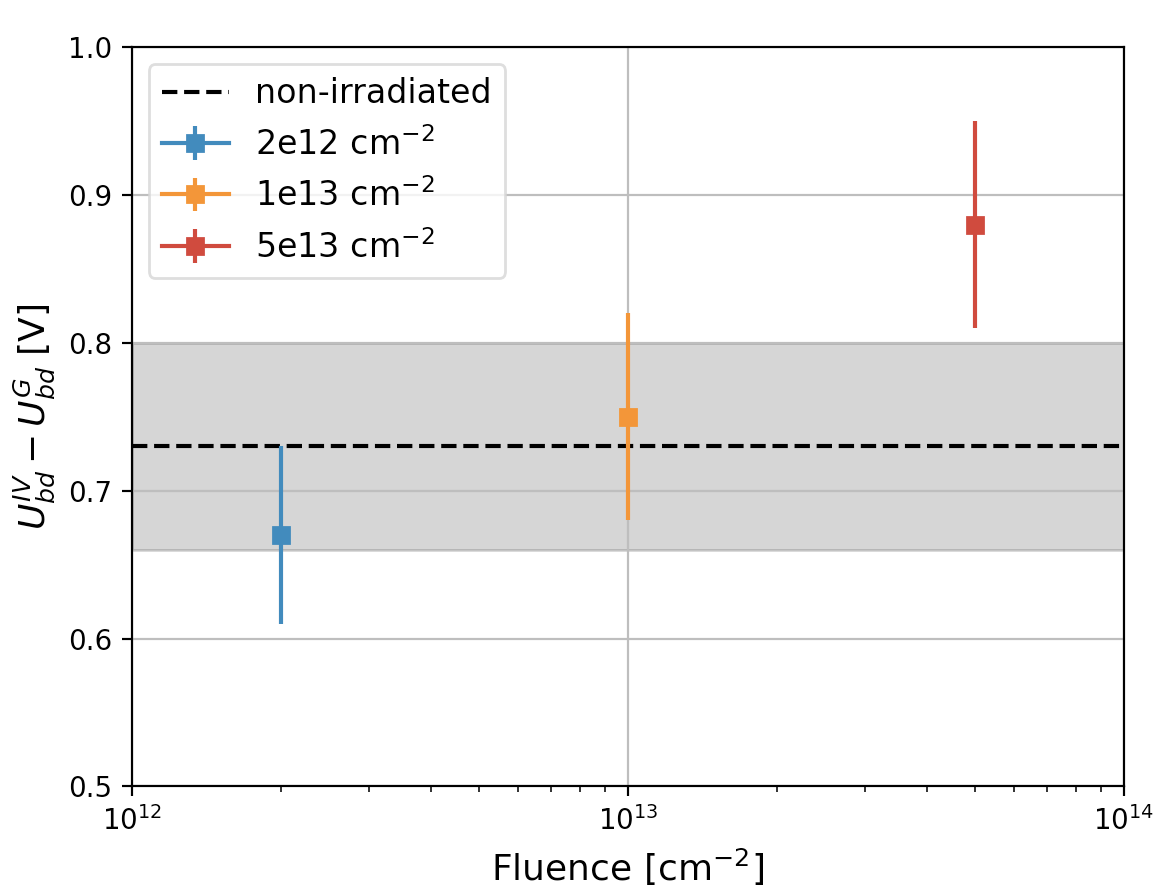}
    \caption{Difference between breakdown voltage from current measurements $U^{IV}_\mathit{bd}$ and breakdown voltage from gain measurements $U^{G}_\mathit{bd}$ as a function of fluence.}
    \label{fig:Vbds_diff}
\end{figure}

Logarithmic derivatives were calculated from the IV-curves measured with illumination for 1 cell and 120 cells. 
The breakdown voltage $U^{IV}_\mathit{bd}$ is determined as a maximum of the logarithmic derivative by approximation with the mean of a Gauss fit. The obtained values of $U^{IV}_\mathit{bd}$ are reported in Table~\ref{table:1}. For all samples $U^{IV}_\mathit{bd}$ of a single cell is equal to $U^{IV}_\mathit{bd}$ of 120 cells within the errors. The table includes also the value of $U^{G}_\mathit{bd}$ reported in~\cite{BYCHKOVA2022166533}. Fig.~\ref{fig:Vbds_diff} presents the answer to the second question posed in the introduction. The difference $U^{IV}_\mathit{bd}-U^{G}_\mathit{bd} \sim $ 0.7~V is approximately contact up to  $\Phi$~=~1e13~cm$^{-2}$, and possibly increases for $\Phi$~=~5e13~cm$^{-2}$. 

\begin{table*}[t]
  \begin{center}
    \begin{tabular}{c|c|c|c} 
      \hline
      \textbf{$\Phi$ [cm$^{-2}$]} & \textbf{Number of cells} & 
      \textbf{U$^{IV}_\mathit{bd}$ [V] at -30~$^{\circ}$C} & \textbf{U$^{G}_\mathit{bd}$ [V] at -30~$^{\circ}$C}\\
      \hline
      \multirow{2}{*}{0e00} & 1 & 35.98$\pm$0.03 & 35.25$\pm$0.04 \\
      & 120 & 35.94$\pm$0.03 & - \\
      \hline
      \multirow{2}{*}{2e12} & 1 & 35.93$\pm$0.02 & 35.26$\pm$0.04 \\
      & 120 & 35.91$\pm$0.03 & - \\
      \hline
      \multirow{2}{*}{1e13} & 1 & 36.16$\pm$0.03 & 35.41$\pm$0.04 \\
      & 120 & 36.13$\pm$0.03 & - \\
      \hline
      \multirow{2}{*}{5e13} & 1 & 36.62$\pm$0.02 & 35.64$\pm$0.06 \\
      & 120 & 36.58$\pm$0.03 & - \\ 
      \hline
    \end{tabular}
  \end{center}
  \caption{Summary of the $U^{IV}_\mathit{bd}$ values for the temperature of -30~$^{\circ}$C.}
  \label{table:1}  
\end{table*}

\begin{comment}
\begin{figure}[h]
  \includegraphics[width=0.9\linewidth]{fig/Vbds_1.png}
  \includegraphics[width=0.9\linewidth]{fig/Vbds_120.png}
  \centering
  \caption{Logarithmic derivatives of a single cell (top) and 120 cells (bottom) for the devices measured at T~=~[-30, +20]~$^{\circ}$C. Data for non-irradiated (black), $\Phi$~=~2e12~cm$^{-2}$ (blue), $\Phi$~=~1e13~cm$^{-2}$ (orange) and $\Phi$~=~5e13~cm$^{-2}$ (red) samples are shown.}
  \label{fig:Vbds}
\end{figure}
\end{comment}

\begin{figure}[h]
  \includegraphics[width=0.9\linewidth]{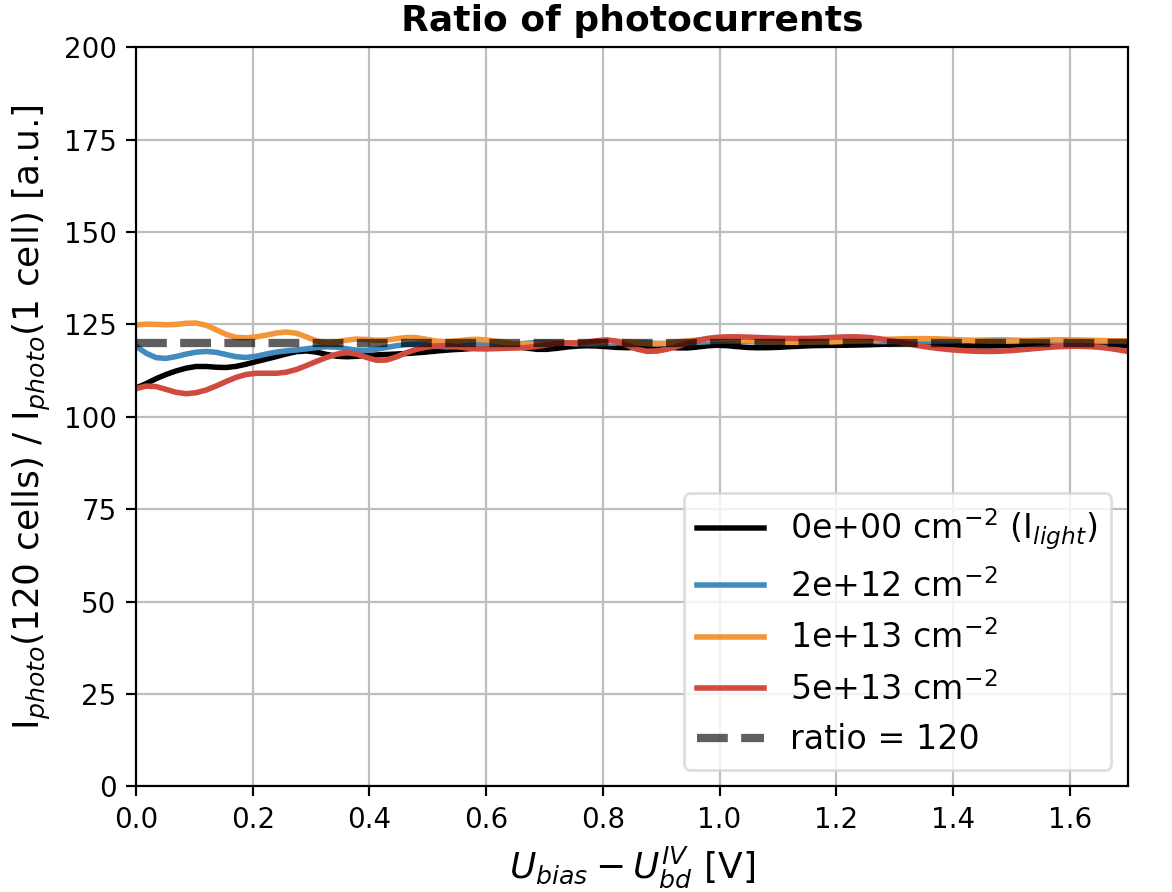}
  \includegraphics[width=0.9\linewidth]{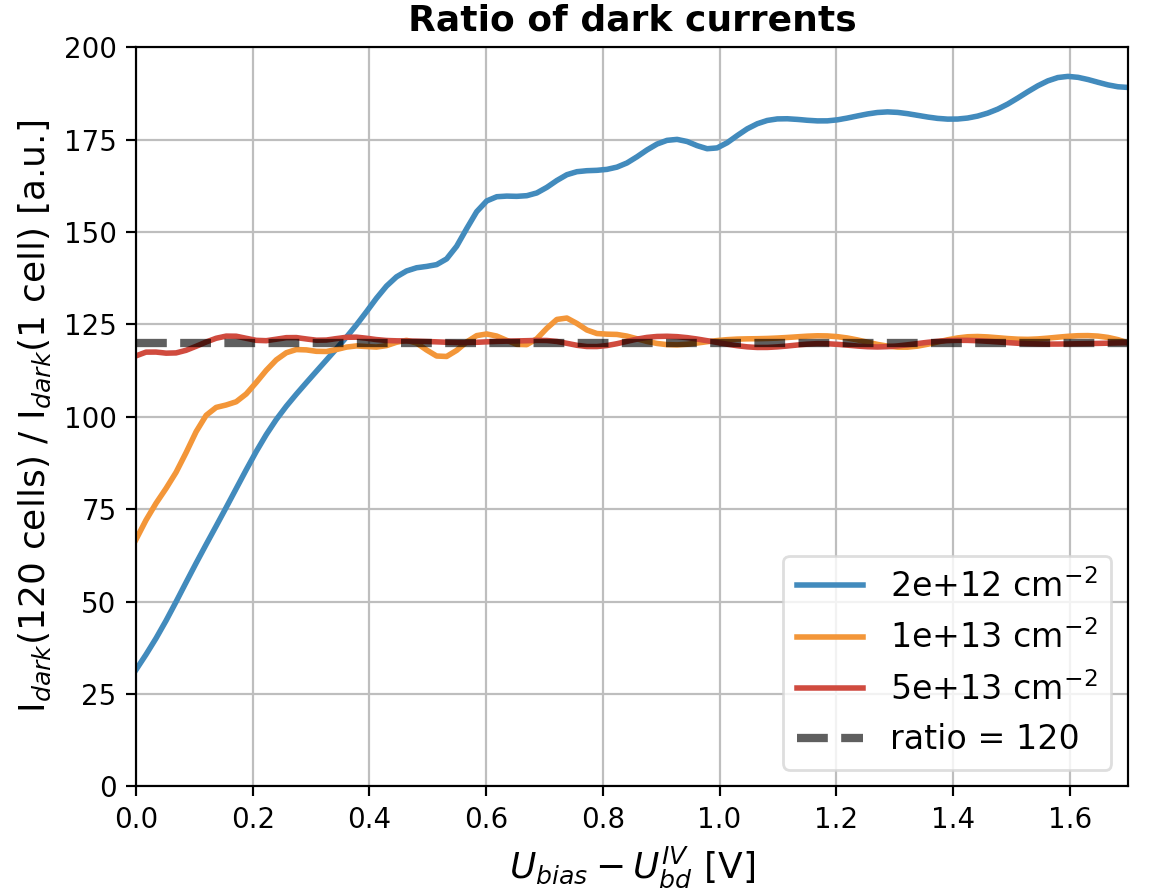}
  \centering
  \caption{Ratio between IV-curves of a single cell and 120 cells for $I_\mathit{photo}$ (top) and $I_\mathit{dark}$ (bottom) measured at T~=~-30~$^{\circ}$C. Data for non-irradiated (black), $\Phi$~=~2e12~cm$^{-2}$ (blue), $\Phi$~=~1e13~cm$^{-2}$ (orange) and $\Phi$~=~5e13~cm$^{-2}$ (red) samples are shown. Expected value of the ratio is marked with a black dashed line. } 
  \label{fig:ratios}
\end{figure}

Radiation damage may produce local "hot spots" with size smaller than or similar to a cell~\cite{Eugen2018:arxiv}.
The probability to find a hot spot in the single cell depends on the fluence and the cell size. Our study is based on the measurements of four single cells, therefore the results could be subject to large
fluctuations due to the presence of "hot spots”. 
To check the uniformity of radiation damage in the SiPM cells, 
we calculate the ratio between the currents of 120 cells and 1 cell for $I_\mathit{dark}$ and $I_\mathit{photo}$. For the non-irradiated sample only the ratio of $I_\mathit{light}$ was calculated, since $I_\mathit{dark}$ is below the accuracy limit of the setup and cannot be measured.
The results are presented in Fig.~\ref{fig:ratios}.

The expected current ratio of 120 is obtained for $I_\mathit{photo}$ at all fluences, for $I_\mathit{light}$ on the non-irradiated sample, and 
for $I_\mathit{dark}$ at the fluences $\Phi$~=~[1e13, 5e13]~$cm^{-2}$ up to $U_\mathit{ov}$~=~1.7~V. 
The ratio for the curves at $\Phi$~=~2e12 $cm^{-2}$ does not reach a plateau and exceeds the value of 120. Possible reasons are: radiation damage non-uniformity or limited accuracy of low current measurements. 
For most of the studied samples the radiation damage of the single
cell is comparable to the average of the surrounding 120 cells,
indicating good damage uniformity both in terms of dark current
change, and of change in the product $PDE \cdot G \cdot(1+CN)$.

\section{Conclusions and outlook}
The radiation hardness study using SiPMs with single-cell readout is ongoing.
First observations from waveform measurements reported a gain reduction by 19\% and an increase of $U^{G}_\mathit{bd}$ by $\approx$0.5~V is observed after $\Phi$~=~5e13~cm$^{-2}$. From the analysis of the IV-curves, the fluence dependence of $U^{IV}_\mathit{bd}$ is extracted, which confirms the same increase of for $U^{G}_\mathit{bd}$, but with a difference in the absolute value of 0.7~V. No visible fluence dependence of the difference $U^{G}_\mathit{bd}-U^{IV}_\mathit{bd}$ is seen within the uncertainties for $\Phi$~=~[2e12, 1e13]~$cm^{-2}$. A light increase of the difference is observed for $\Phi$~=~5e13~$cm^{-2}$.
The radiation damage uniformity of 1 cell and 120 cells was checked up to $U_\mathit{ov}$~=~1.7~V. A good damage uniformity both in terms of dark- and photocurrent change is confirmed on all but one sample at the lowers investigated fluence. 

With a dedicated study on non-irradiated sensors it is shown that self-heating effects are negligible for the power dissipated even in the highest irradiated SiPM.

\section*{Acknowledgement}
This work is supported by the Deutsche Forschungsgemeinschaft (DFG, German Research Foundation) under Germany's Excellence Strategy, EXC 2121, Quantum Universe (390833306).
The reported study was funded by RFBR and TUBITAK according to the research project 20-52-46005.

\bibliographystyle{elsarticle-num}
\bibliography{article}

\end{document}